\documentclass[fleqn,12pt,twoside]{article}
\usepackage{espcrc1}

\def\bq{\begin{eqnarray}}
\def\eq{\end{eqnarray}}
\def\be{\begin{eqnarray}}
\def\ee{\end{eqnarray}}

\usepackage{epsfig}

\newcommand{\AmS}{{\protect\the\textfont2
  A\kern-.1667em\lower.5ex\hbox{M}\kern-.125emS}}

\title{
The effects of nuclear structure on generalized parton distributions
of $^3$He
}

\author{S. Scopetta\address[MCSD]{
Dipartimento di Fisica, Universit\`a degli Studi
di Perugia, via A. Pascoli, 06100 Perugia, Italy
and INFN, sezione di Perugia}}
       
\begin{document}

\maketitle

\begin{abstract}
The effect of the nuclear medium on
generalized parton distributions (GPDs) is studied
for the $^3$He nucleus,
through a realistic microscopic analysis.
In Impulse Approximation, 
Fermi motion 
and binding effects, evaluated by 
modern potentials, 
are found to be larger than in the forward case
and
%
very sensitive to the details of nuclear structure at short distances.
\end{abstract}

\vskip 0.5cm


Generalized Parton Distributions (GPDs) \cite{first} 
enter the long-distance dominated part of
exclusive lepton Deep Inelastic Scattering
(DIS) off hadrons
(for recent reviews, 
see, e.g., Ref. \cite{dpr}).
Deeply Virtual Compton Scattering (DVCS),
i.e. the process
$
e H \longrightarrow e' H' \gamma
$ when
$Q^2 \gg m_H^2$, permits to access GPDs
(here and in the following,
$Q^2$ is the momentum transfer between the leptons $e$ and $e'$,
and $\Delta^2$ the one between the hadrons $H$ and $H'$) so that
relevant experimental DVCS programs are taking place.
Presently, the issue of measuring GPDs for nuclei
is being addressed \cite{bar}. 
As observed already in \cite{cano1}, 
the knowledge of GPDs would permit the investigation
of the short light-like distance structure of nuclei, and thus the interplay
of nucleon and parton degrees of freedom in the  
wave function.
In DIS off a nucleus
with four-momentum $P_A$ and $A$ nucleons of mass $M$,
this information can be accessed in the 
region where
$A x_{Bj} \simeq { Q^2 / ( 2 M \nu) }>1$,
being $x_{Bj}= Q^2 / ( 2 P_A \cdot q )$ and $\nu$
the energy transfer in the laboratory system.
In this region measurements are difficult, because of 
vanishing cross-sections.
The same physics can be accessed
in DVCS at lower values of $x_{Bj}$ 
\cite{cano2}. 
The study of GPDs for $^3$He is interesting
for many aspects. 
$^3$He is a well known nucleus, for which realistic studies 
are possible, so that conventional effects
can be calculated and the exotic ones
can be investigated.
Besides, $^3$He is extensively used as an effective 
polarized neutron target \cite{friar} and
it will be the first candidate
for experiments aimed at the study of
GPDs of the free neutron, to unveil its angular momentum
content. 
In this talk, 
the results of an impulse approximation (IA)
calculation \cite{prc} of the quark unpolarized GPD $H_q^3$ of
$^3$He are reviewed. A convolution formula
is discussed and evaluated
using a realistic non-diagonal spectral function,
so that Fermi motion and binding effects are rigorously estimated.
The proposed scheme is valid for 
$\Delta^2 \ll Q^2,M^2$
and despite of this it permits to
calculate GPDs in the kinematical range relevant to
the coherent channel of DVCS off $^3$He.
In fact, the latter channel is the most interesting one for its 
theoretical implications, but it can be hardly observed at
large $\Delta^2$, due to the vanishing cross section.
This investigation
permits to test
the accuracy of prescriptions
which have been proposed to estimate nuclear GPDs \cite{cano2}.


If one thinks to a spin $1/2$ hadron target, with initial (final)
momentum and helicity $P(P')$ and $s(s')$, 
respectively, two 
GPDs $H_q(x,\xi,\Delta^2)$ and
$E_q(x,\xi,\Delta^2)$, occur.
If one works in a system of coordinates where
the photon 4-momentum, $q^\mu=(q_0,\vec q)$, and $\bar P=(P+P')/2$ 
are collinear along $z$,
$\xi$ is the so called ``skewedness'', parametrizing
the asymmetry of the process, defined
by the relation 
$
\xi = - {n \cdot \Delta / 2} = - {\Delta^+ / 2 \bar P^+}
= { x_{Bj} /( 2 - x_{Bj}) } + {{O}} \left ( {\Delta^2 / Q^2}
\right ) ~,
$
where $n$
is a light-like 4-vector
satisfying the condition $n \cdot \bar P = 1$.
One should notice that the variable $\xi$
is completely fixed by the external lepton kinematics.
The well known constraints of $H_q(x,\xi,\Delta^2)$ are: 
i) the so called
``forward'' limit, 
$P^\prime=P$, i.e., $\Delta^2=\xi=0$, where one 
recovers the usual PDFs
$
H_q(x,0,0)=q(x)~;
\label{i)}
$
ii)
the integration over $x$, yielding the contribution
of the quark of flavor $q$ to the Dirac 
form factor (f.f.) of the target:
$
\int dx H_q(x,\xi,\Delta^2) = F_1^q(\Delta^2)~;
\label{ii)}
$
iii) the polynomiality property.

In Ref. \cite{prc}, specifying to the $^3$He target
the procedure developed in
Ref. \cite{io3},
an IA expression
for $H_q(x,\xi,\Delta^2)$ of a given hadron target
has been obtained.
Assuming that the interacting parton belongs
to a bound nucleon with momentum $p$ and removal energy $E$,
for small values of $\xi^2$ and
$\Delta^2 \ll Q^2,M^2$, it reads:

\begin{eqnarray}
\label{flux}
H_q^3(x,\xi,\Delta^2) 
& =  & 
\sum_N \int dE \int d \vec p
\, 
P_{N}^3(\vec p, \vec p + \vec \Delta, E )
{\xi' \over \xi}
H_{q}^N(x',\xi',\Delta^2)~.
\label{spec}
\end{eqnarray}
In the above equation, the kinetic energies of the residual nuclear
system and of the recoiling nucleus have been neglected, 
$P_{N}^3 (\vec p, \vec p + \vec \Delta, E )$ is
the one-body off-diagonal spectral function
for the nucleon $N$ in $^3$He,
the quantity
$
H_q^N(x',\xi',\Delta^2)
$
is the GPD of the bound nucleon N
up to terms of order $O(\xi^2)$, and in the above equation
use has been made of
the relations
$
\xi'  =  - \Delta^+ / 2 \bar p^+~,
$
and $ x' = (\xi' / \xi) x$~.
Eq. (\ref{spec}) can be written in the form
\begin{eqnarray}
H_{q}^3(x,\xi,\Delta^2) =  
\sum_N \int_x^1 { dz \over z}
h_N^3(z, \xi ,\Delta^2 ) 
H_q^N \left( {x \over z},
{\xi \over z},\Delta^2 \right)~,
\label{main}
\end{eqnarray}
where 
$ h_N^3(z, \xi ,\Delta^2 ) =  
\int d E
\int d \vec p
\, P_N^3(\vec p, \vec p + \vec \Delta) 
\delta \left( z + \xi  - { p^+ / \bar P^+ } \right)~.
$

In Ref. \cite{prc}, it is discussed that
Eq. (\ref{spec}) fulfills the constraints $i)-iii)$ previously listed.


$H_q^3(x,\xi,\Delta^2)$, Eq. (\ref{spec}), 
has been evaluated in the nuclear Breit Frame.
The non-diagonal spectral function appearing in
Eq. (\ref{spec}) has been calculated 
along the lines of Ref. \cite{gema},
by means of 
realistic wave functions 
evaluated
using the 
AV18 interaction and
taking into account
the Coulomb repulsion.
The other ingredient in Eq. (\ref{spec}), i.e.
the nucleon GPD $H_q^N$, has been modelled in agreement with
the Double Distribution representation \cite{rad1}.
In this model, whose details are summarized in Ref. \cite{prc},
the $\Delta^2$-dependence of $H_q^N$ is given by
$F_q(\Delta^2)$, i.e. the contribution
of the quark of flavor $q$
to the nucleon form factor.
Now the numerical results
will be presented.
If one considers
the forward limit of the ratio
\bq
R_q (x,\xi,\Delta^2) = 
{ H_q^3(x,\xi,\Delta^2) 
/ ( 2 H_q^p(x,\xi,\Delta^2) + H_q^n(x,\xi,\Delta^2) )}~,
\label{rat}
\eq
where the denominator clearly represents
the distribution of the
quarks of flavor $q$ 
in $^3$He if nuclear effects are completely
disregarded, 
the behaviour which is found, shown in Ref. \cite{prc},
is typically $EMC-$like,
so that, 
in the forward limit, well-known results are recovered.
In Ref. \cite{prc} it is also shown that
the $x$ integral of the nuclear GPD gives a good 
description of ff data of $^3$He, in the relevant kinematical region,
$-\Delta^2 \leq 0.25$ GeV$^2$.
\begin{figure}[ht]
\centerline{\epsfxsize=2.5in\epsfbox{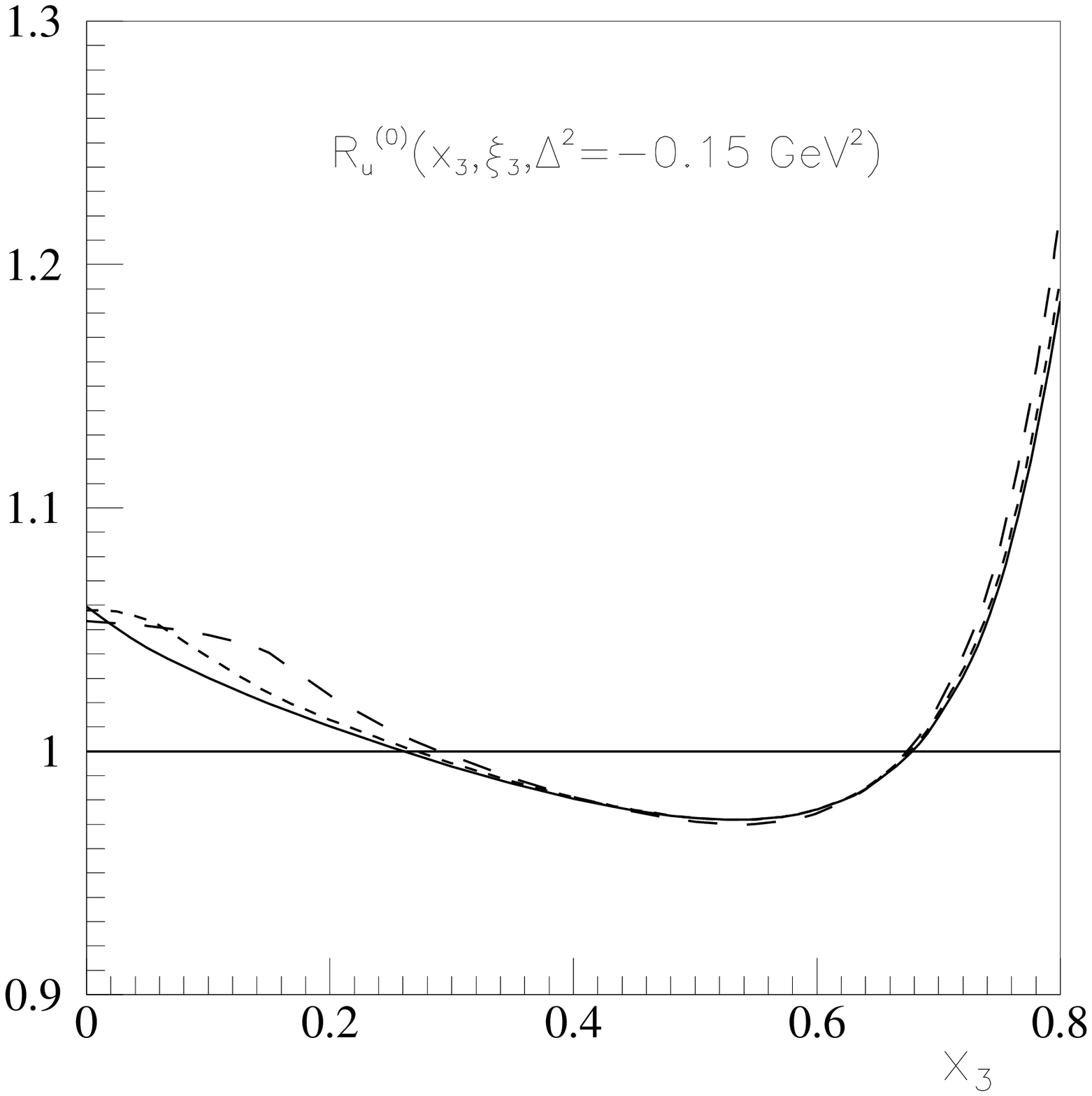}
\epsfxsize=2.5in\epsfbox{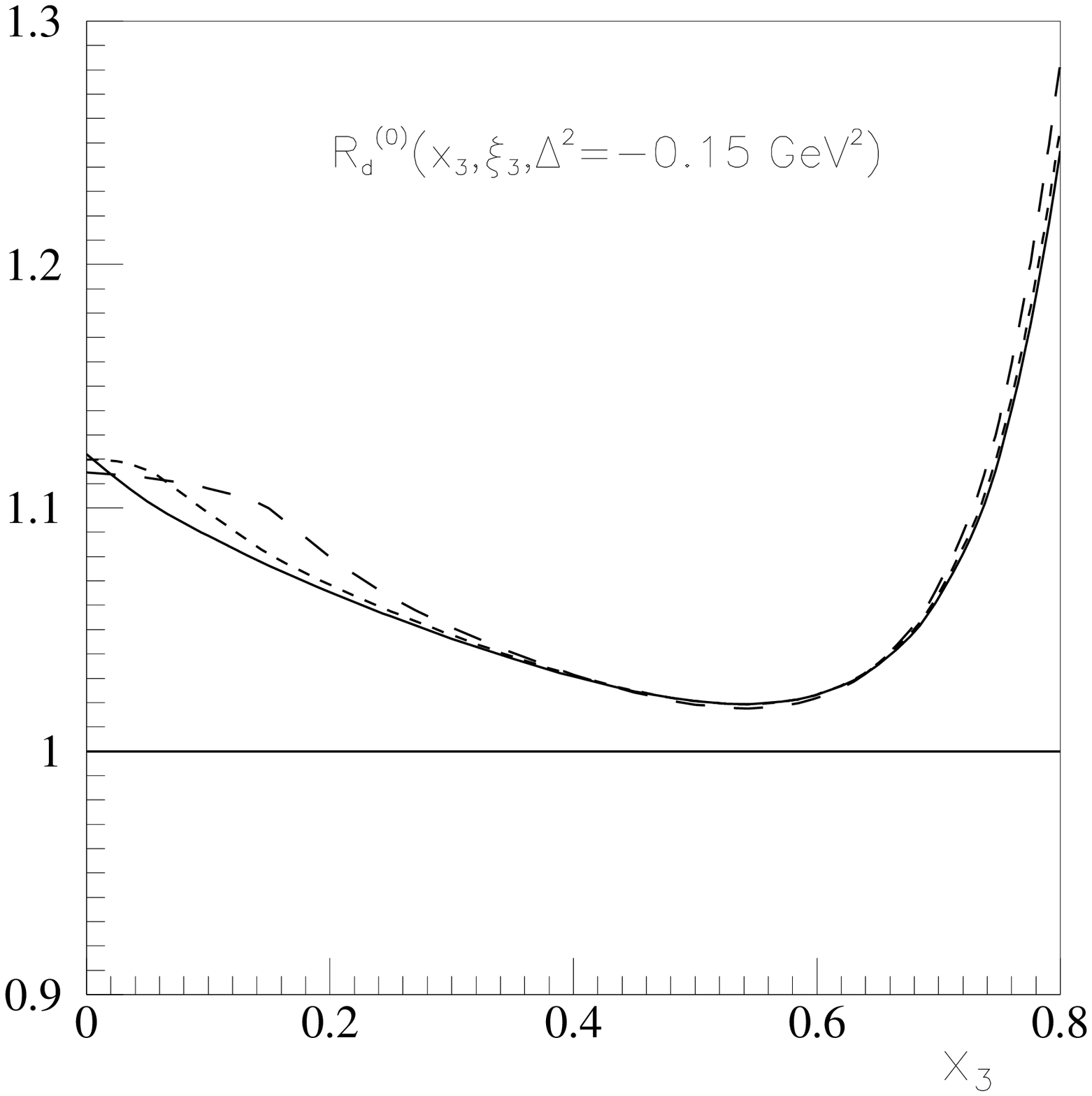}}   
\caption{
In the left panel,
the ratio Eq. (\ref{rnew}) is shown, for the $u$ flavor and
$\Delta^2 = -0.15$ GeV$^2$, 
as a function of $x_3$.
The full line has been calculated for $\xi_3=0$,
the dashed line for $\xi_3=0.1$ and the long-dashed
one for $\xi_3=0.2$. 
The symmetric part at $ x_3 \leq 0$ is not presented.
In the right panel, the same is shown, for the flavor $d$.}
\end{figure}
Let us now discuss the nuclear effects.
The full result for the GPD $H_q^3$, Eq. (\ref{spec}),
will be compared with a prescription
based on the assumptions
that nuclear effects are neglected
and the $\Delta^2$ dependence can be
described 
by
the f.f. of $^3$He:
\bq
H_q^{3,(0)}(x,\xi,\Delta^2) 
= 2 H_q^{3,p}(x,\xi,\Delta^2) + H_q^{3,n}(x,\xi,\Delta^2)~,
\label{app0}
\eq
where the quantity
$
H_q^{3,N}(x,\xi,\Delta^2)=  
\tilde H_q^N(x,\xi)
F_q^3 (\Delta^2)
$
represents the flavor $q$ effective GPD of the bound nucleon 
$N=n,p$ in $^3$He. Its $x$ and $\xi$ dependences, given by the function
$\tilde H_q^N(x,\xi)$, 
is the same of the GPD of the free nucleon $N$,
while its $\Delta^2$ dependence is governed by the
contribution of the quark of flavor $q$ to the
$^3$He f.f., $F_q^3(\Delta^2)$.

The effect of Fermi motion
and binding can be shown through 
the ratio
\be
R_q^{(0)}(x,\xi,\Delta^2) = { H_q^3(x,\xi,\Delta^2) / H_q^{3,(0)}
(x,\xi,\Delta^2)} 
\label{rnew}
\eq
i.e. the ratio
of the full result, Eq. (\ref{spec}),
to the approximation Eq. (\ref{app0}).
\begin{figure}[ht]
\centerline{\epsfxsize=2.5in\epsfbox{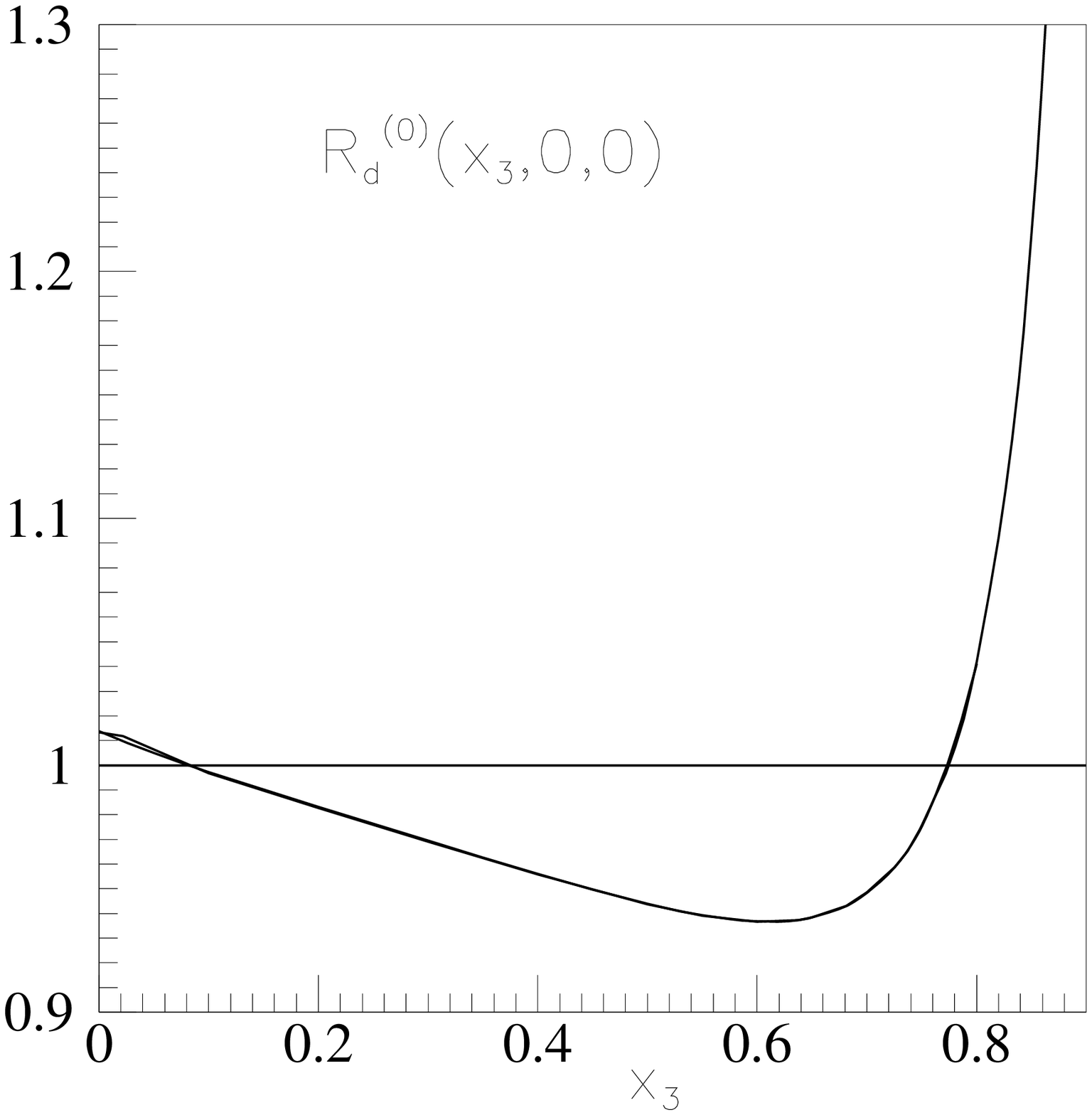}
\epsfxsize=2.5in\epsfbox{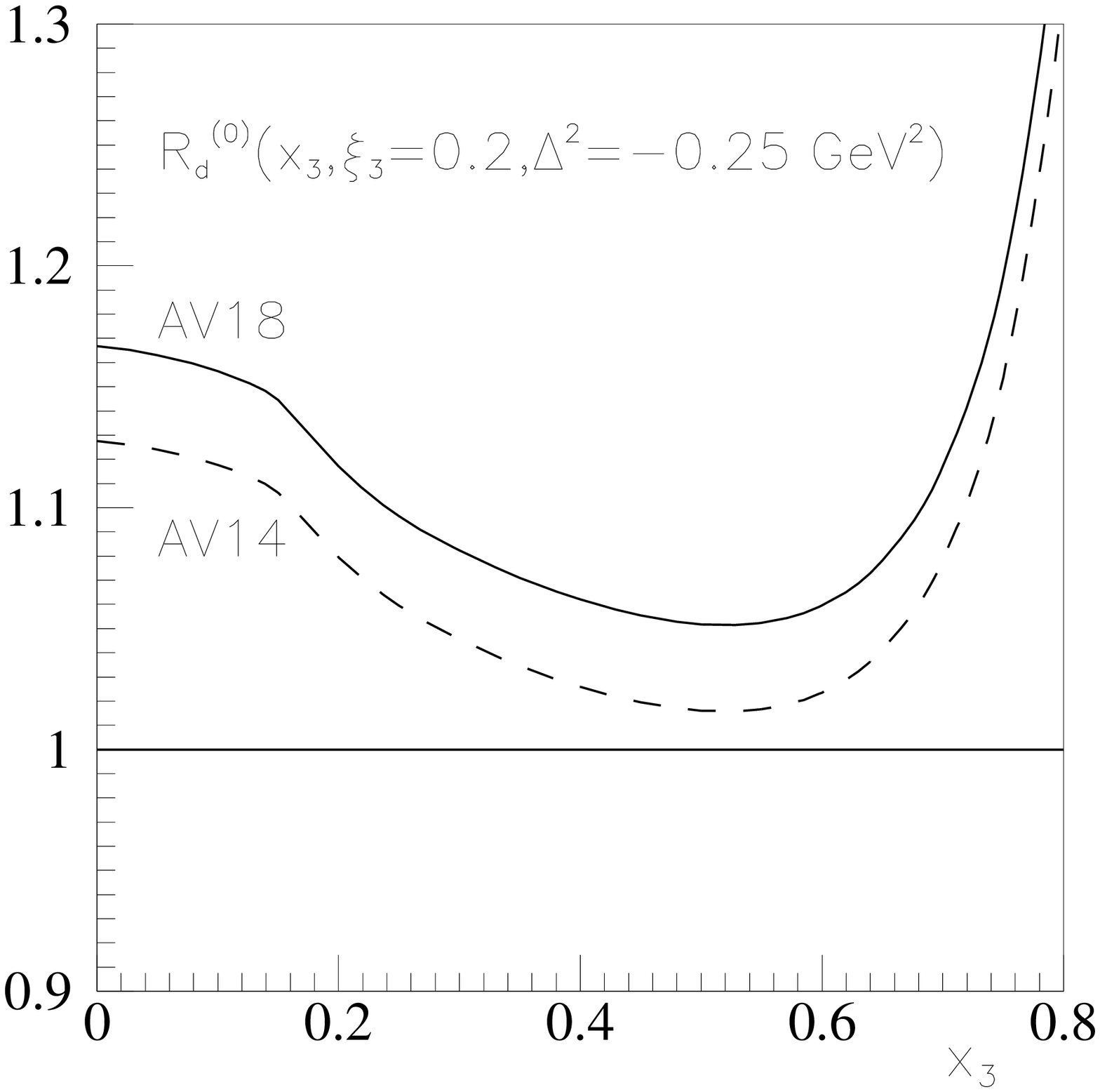}}   
\caption{
Left panel: the ratio $R^{(0)}$, for the
$d$ flavor, in the forward limit $\Delta^2 = 0, \xi=0$, calculated
by means of the AV18 (full line) and AV14
(dashed line) interactions, as 
a function of $x_3 = 3 x$.
The results obtained with the different potentials are
not distinguishable.
Right panel: the same
as in the left panel, but at $\Delta^2=-0.25$ Ge$V^2$ 
and $\xi_3 = 3 \xi = 0.2$. The results are now
clearly distinguishable.
}
\end{figure}
The choice of calculating the ratio Eq. (\ref{rnew})
to show nuclear effects is a very natural one.
As a matter of fact, the forward limit of the ratio Eq. (\ref{rnew})
is the same of the ratio Eq. (\ref{rat}), yielding the
EMC-like ratio for the parton distribution $q$ and,
if $^3$He were made of free nucleons at rest,
the ratio Eq. (\ref{rnew}) would be one.
This latter fact can be immediately realized by
observing that the prescription Eq. (\ref{app0})
is exactly obtained by
placing $z=1$, i.e. no Fermi motion effects
and no convolution, into Eq. (\ref{main}). 
Results are presented in Fig. 1, where
the ratio Eq. (\ref{rnew}) is shown
for $\Delta^2 = -0.15$ GeV$^2$ 
as a function of $x_3=3 x$,  
for three different values
of $\xi_3=3 \xi$, for the flavors $u$ and $d$.
Some general trends of the results are apparent:
i) nuclear effects, for $x_3 \leq 0.7$, are as large as 15 \% at most;
ii) Fermi motion and binding have their main effect
for $x_3 \leq 0.3$, at variance with what happens
in the forward limit;
iii) nuclear effects increase with
increasing $\xi$ and
$\Delta^2$, for $x_3 \leq 0.3$;
iv) nuclear effects for the $d$ flavor are larger than
for the $u$ flavor.
The behaviour described above is discussed and explained
in Ref. \cite{prc}.
In Fig. 2, it is shown that nuclear effects
are found to 
depend also on the choice of the NN potential, 
at variance with what happens in the forward case \cite{gron}.
Nuclear GPDs turn out therefore to be strongly dependent on
the details of nuclear structure.
The obtained GPDs are being used to 
estimate cross-sections and to establish
the feasibility of experiments. 
The study of polarized GPDs
will be very interesting, due to its implications
for unveiling the angular momentum of the free neutron.

\end{document}